\documentclass [12pt] {article}

\begin{document}

\title{Proposal:  The Neural Network Telescope}
\author{{\.{I}brahim Semiz \thanks{mail: ibrahim.semiz@boun.edu.tr}}\\    
   Department of Physics  \\
   Bo\u{g}azi\c{c}i University\\
Bebek, \.{I}stanbul, TURKEY\\}

\maketitle

\begin{abstract}
A neural network mechanism that can compensate for poor optical quality was recently discovered in a biological context. We propose that this mechanism can and should be adopted for astronomical purposes. This would shift emphasis away from the quality of the optical equipment to information processing, hence should decrease the cost and make larger instruments feasible. 
\end{abstract}

In astronomy, it is desired that a telescope be as large as possible, so that
light-gathering power is increased and fainter sources can be seen. However, producing large mirrors to the accuracy desired for sharp image formation at the focus is very technically challenging and prohibitively expensive. Also, large mirrors are very massive, and their weight puts considerable mechanical stress on the support structure and can deform the mirror itself. Therefore there exist no single-mirror telescopes with aperture larger than 8 m.

Here we propose another method for astronomical observation: We do not require the telescope to focus the light \emph{exactly}, but only to collect it onto the pixellated light detector ("PLD", e.g. a CCD). Then a neural network would recreate the image by taking superpositions of all PLD pixel signals for all image pixels.

Nature already does this, as reported in a recent Letter  \cite{viper}, whose results we summarize briefly: 

Some snakes possess an infrared sense organ ("IR pit") which is an optically very poor camera, yet they are able to achieve  good spatial resolution. This is accomplished by a network of synaptic connections between the infrared-sensing cells in the IR pit and the \emph{optic tectum} in the brain. This neural network has to be trained by comparing the optic image from the eye with the "image" from the IR pit \cite{pt}.

Consider an object to made up of equal number of points as the infrared-sensing cells (and in the same geometrical configuration). Each sensing cell may receive light from each object point. If the object points and the sensing cells are labeled by $i$ and $j$ respectively, and the intensity emitted and detected by the image points and sensing cells are denoted by $x_i$ and $y_j$ respectively, the effect of the IR pit can be described by a transformation matrix:

\begin{equation}
\vec{y} = T_{1} \vec{x}			\label{trf}
\end{equation}
where the matrix $T_{1}$ is determined by the geometry of the IR pit. 

Now, consider the intensities received by the sensing cells to be recombined again to give the intensities of another set of points $z_k$, which we can call "image" pixels:

\begin{equation}
\vec{z} = T_{2} \vec{y}			\label{rec}
\end{equation}
Here, the elements of the matrix $T_{2}$ are the strength of the synaptic connections between the $ j^{\rm th}$ sensing cell and $k^{\rm th}$ region in the  \emph{optic tectum}. Of course, successful  IR vision means that the image is the same as the object, i.e. $ \vec{z}=\vec{x}$, therefore it must be arranged for the matrix $T_{2}$ to be the inverse of the matrix $T_{1}$. The training of the IR vision of the snake consists of the adjustments of the connection strengths by inhibitory and excitatory neural interactions so that this is the case.

The \emph{neural network telescope} ("NNT") we propose in this Letter embodies the same principle. A large light-collecting device, possibly a mirror, takes the place of the IR pit, a human-made PLD the place of the sensing cells, a hardwired neural network or a computer the place of the synaptic connections, and a display the place of the  \emph{optic tectum}.

To "train" the NNT, we must determine the matrix elements of $T_{2}$. To this end, we can point the telescope to a (real or artificial) point source in its field of view so that only one $x_i$ is nonzero, measure all $y_i$, shift the telescope (or the artificial point source) slightly so that the next $x_i$ is nonzero and repeat, thereby determine the matrix elements of $T_{1}$ column by column. Inverting it will give the desired $T_{2}$, which then can be built in as strength of the corresponding connections or used by the reconstruction software.

Of course, in a perfect regular telescope, the focusing of light by the mirror makes a one-to-one assignment between the object points and the PLD pixels, i.e. $T_{1}$ is proportional to the identity matrix.

The snakes investigated in \cite{viper} have about 40x40 sensing cells, while a PLD may be of the order of megapixels (the original HST camera was 800x800 pixels \cite{hubblepixels}). Since the inversion of an $N \times N$ matrix takes $O(N^{2.376})$ operations \cite{copp-wino}, we need about $10^{5}$ s on a 10 GigaFLOPS computer for a one-megapixel sensor. The standard desktop computers are approaching this performance, and $10^{5}$ s is acceptable, since the inversion needs to be done only once. Image reconstruction corresponds to the multiplication (\ref{rec}), which takes  $O(N^{2})$ operations and therefore can be done in a few hundred seconds. Of course, dedicated computers will be faster, in fact, some consumer game consoles and graphic cards claim speeds of TeraFLOPS; naturally, future computers will be even faster.

The angular resolution of a NNT will be
\begin{equation}
R \sim \frac{\Delta\Theta}{n}
\end{equation}
where $\Delta\Theta$ is the angular size of the field of view, and $n$ is the number of pixels on one edge of the PLD ($N \sim n^{2}$). In the snakes' case, the expected resolution is $100^{\circ} / 40 = 2.5^{\circ}$  \cite{viper}; of course in astronomy, much better resolution is needed. The obvious way of increasing resolution for a given PLD is making the field of view smaller; in fact, it can be said that a regular telescope does just this: The mirror makes the field of view of a given pixel almost zero.

Therefore the optical, that is, light-gathering component of the NNT is required to collect --but not necessarily focus-- light from a  \emph{narrow field of view} onto the PLD, and prevent light from outside  the desired field of view from reaching it. Since the focusing properties do not have to be very precise, the light-gathering component can presumably be made cheaper and/or bigger then for regular telescopes. For example, 

(a) Monolithic mirrors comparable in size to the largest existing ones can be made at a fraction of their cost and in shorter time.

(b) Arbitrarily large segmented mirrors can be constructed with {\em flat segments}, significantly reducing cost and production time; and by not attaching segments rigidly to each other, and controlling their position and orientation independently, thermal and gravitational deformations can be counteracted.

\begin{figure}

\caption{The "pipectack" option for the NNT. PLD is the grid at the closer end. \label{pipestack}}

\begin{picture}(400, 400)

\multiput(0,0)(0,5){21}{\line(1,0){100} }
\multiput(0,0)(5,0){21}{\line(0,1){100} }

\multiput(3,103)(10,0){10}{\line(1,1){200} }
\multiput(103,3)(0,10){10}{\line(1,1){200} }

\multiput(6,99)(10,0){9}{\oval(10,10)[t]}
\multiput(99,6)(0,10){9}{\oval(10,10)[r]}
\put(99,99){\oval(10,10)[r]}
\put(99,99){\oval(10,10)[t]}

\multiput(206,299)(10,0){9}{\oval(10,10)[t]}
\multiput(299,206)(0,10){9}{\oval(10,10)[r]}
\put(298,298){\oval(10,10)[r]}
\put(298,298){\oval(10,10)[t]}

\put(300,350){to object}

\put(85,250){\large pipestack}

\put(160,30){\large PLD}

\thicklines
\qbezier(50,50)(100,50)(155,35)
\put(55,50){\vector(-1, 0 ){5}}
\put(330,330){\vector(1, 1 ){50}}

\end{picture} 
\end{figure}
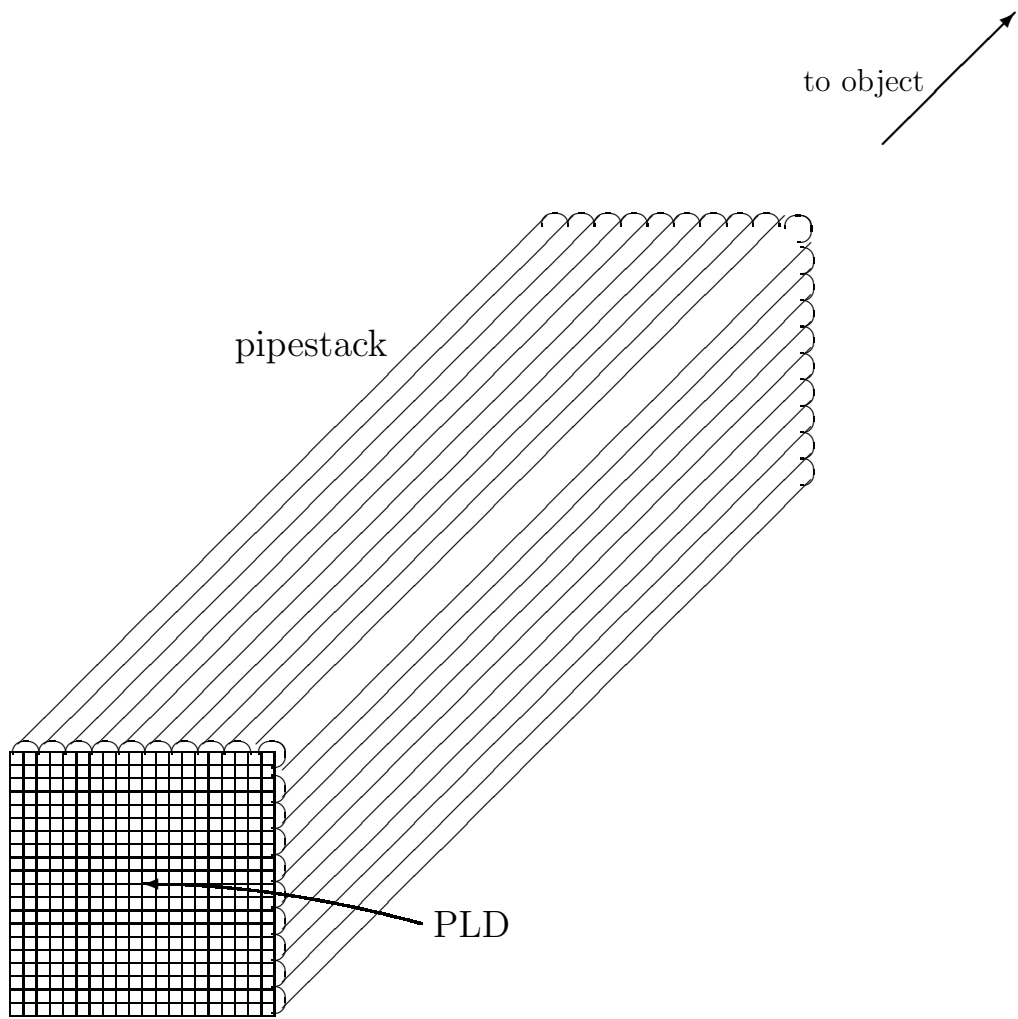

(c) In a design with no mirrors whatsoever, a stack of parallel pipes with light-absorbing insides (Figure \ref{pipestack}) can be used to limit the field of view, which would then be the ratio of the width to length of a pipe.

The area onto which light is collected may be large compared to regular telescopes, because of imprecise focus in (a) and (b), and no focus in (c). In this case, an array of photomultiplier tubes or avalanche photodiodes may also serve as the PLD, although this may drive the cost up with current prices.

All of the above options are suitable for space telescopes, as well. The parts of both the very large telescope in (b) and the pipestack in (c) can be mass-produced on Earth or in space, since high precision is not needed; and the telescope assembled and calibrated (trained) on location. Of course, going to space also has the advantage of eliminating gravitational deformations.

Unlike focusing accuracy, the rigidity of the optical element is essential. Gravitational or thermal deformations will change the matrix $T_{1}$, making the (real or virtual) neural network unusable. In option (b) above, this problem can, and should be, corrected. In some cases, if the temperature of the optical element is known to be uniform across its structure, but changes slowly in time, it might be possible to make the matrices $T_{1}$ and $T_{2}$ functions of temperature.

Color images can be produced by superposing images taken with three color filters, but spectroscopy with the NNT seems difficult or impossible, since it does not produce a narrow collimated beam which can be diffracted by a grating. This may change if future technology can produce pixels capable of measuring photon energies to high accuracy.

Another advantage of the NNT is that it can work in any part of the spectrum, including X- or $\gamma$-rays, which cannot be focused in the usual, optical sense of the word. It is also interesting to note that the NNT idea would make no sense when the human eye or photographic film was the light-detector; it only would work in the CCD era.

We would like to thank T. Rador and E. G\"{u}lmez for helpful discussions. This work was partially supported by grant 06B303 of Bo\~gazi\c{c}i University Research fund.

\end{document}